\documentclass[
    ,final            
  ]
  {aipproc}

\layoutstyle{6x9}
\usepackage{epsfig,color,axodraw}
\begin{document}


\title{Precision Observables in the MSSM: W mass
 and the muon magnetic
moment
}


\author{J.\ Haestier}{
  address={IPPP, University of Durham, Durham, UK}
}

\author{S.\ Heinemeyer}{
  address={Instituto de Fisica de Cantabria (CSIC-UC), Santander, Spain}
}

\author{W.\ Hollik}{
  address={Max-Planck-Institut f\"ur Physik, Munich, Germany}
}

\author{{\underline{D.\ St\"ockinger}} 
\footnote{Talk given by D.S.\ at the SUSY06 Conference  on Supersymmetry
and the Unification of Fundamental Interactions
Irvine, California, USA, 12--17 June 2006}
}{
  address={SUPA, School of Physics, Edinburgh University, UK}
}

\author{\newline A.\ M.\ Weber}{
  address={Max-Planck-Institut f\"ur Physik, Munich, Germany}
}

\author{G.\ Weiglein}{
  address={IPPP, University of Durham, Durham, UK}
}

\begin{abstract}
The precision observables $M_W$ and $g-2$ of the muon are discussed
in the framework of the MSSM. Recent progress in the evaluation of the 
theoretical predictions is described, and the MSSM predictions are compared 
with the SM predictions and the experimental values.
\end{abstract}

\maketitle


\section{Introduction}

Precision observables are a unique laboratory to test the Standard Model (SM) 
or extensions at the quantum level. Via quantum effects, heavy particles enter 
the theoretical predictions for such observables, and comparing
measurements with predictions leads to valuable information e.g.\ on the masses
of postulated particles such as Higgs bosons or supersymmetric
particles.

The power of precision observables is illustrated by comparing the current
experimental resolution to the numerical size of quantum effects in the SM
and the minimal supersymmetric standard model (MSSM). 
In the case of the mass of the $W$ boson, $M_W$,
the  SM one-loop and two-loop effects
amout to about $(-15,-3)$ times the current experimental
uncertainty of $0.04\%$
\cite{EWPOexp}.
The weak SM one-loop and two-loop contributions to the anomalous 
magnetic moment of the muon $a_\mu=(g-2)_\mu/2$
 are about $(4,-1)$ times as large
as  the current experimental uncertainty of $0.54$ parts per million
\cite{BNL56}.

The MSSM is a weakly coupled,
renormalizable gauge theory \cite{MSSMRen}, and therefore quantum effects
are well-defined and calculable.
In the MSSM, quantum effects
from supersymmetric (SUSY) particles to $M_W$ and $a_\mu$ depend on many MSSM 
parameters, but
they can be as large as the corresponding
SM quantum effects and thus significantly larger than the 
experimental uncertainty. Conversely,
the experimental measurements significantly constrain the
SUSY parameter space (see e.g.\ the reviews \cite{PomssmRep,rev}).

In these proceedings we give an update on
the current status and recent theoretical 
developments of the two observables $M_W$ and $a_\mu$ in the MSSM.

\section{$M_W$ in the MSSM}

The mass of the $W$-boson $M_W$ has been measured at LEP and is being 
measured at Tevatron. The current experimental value is $M_W^{\rm exp}=
80.392(29)$ \cite{EWPOexp}, and the precision could be improved to
 $\delta_{\rm LHC}M_W=15$~MeV at the LHC \cite{mwlhc}
and to $\delta_{\rm ILC}M_W=7$~MeV  at a linear $e^+e^-$ collider 
\cite{mwgigaz}.
 On the theoretical side, the SM or MSSM predict a calculable
relation between $M_W$ and the muon lifetime $\tau_\mu$ and $M_Z$. Solving
this relation for $M_W$ leads to a prediction of $M_W$ in terms
of $\tau_\mu$, $M_Z$ and all other model
parameters, in particular the masses of the top quark, Higgs bosons, and 
SUSY particles in the case of the MSSM.
For the SM prediction of $M_W$ see  \cite{MWSM}
and
references therein; for a review of previously available MSSM contributions
to $M_W$ see \cite{PomssmRep}.

More recently, the Yukawa-enhanced
 ${\cal O}(\alpha_t^2,\alpha_t\alpha_b,\alpha_b^2)$ contributions to 
$M_W$
have been evaluated \cite{drMSSMal2B}. 
This result completes the evaluation
of all two-loop MSSM contributions to $M_W$ that enter via the quantity
$\Delta\rho$.
Detailed estimates for the remaining two-loop contributions, 
which go beyond 
$\Delta\rho$, and for unknown higher-order contributions have been derived
 \cite{drMSSMal2B}.

In \cite{MWMSSM}, all existing SM and MSSM results have been combined with a
new evaluation of the one-loop MSSM contribution that also takes into
account complex phases. In this way, a very precise and reliable prediction
for $M_W$ in the MSSM has been obtained. This prediction has
been implemented in a computer code that will be made
publicly avaliable. The remaining theory error of this MSSM 
prediction of $M_W$, due to the unknown multi-loop contributions estimated in
\cite{drMSSMal2B} and to the unknown phase dependence beyond the one-loop
level \cite{MWMSSM},
has been estimated to $\delta M_W=(4.7 - 10.6)$~MeV,
depending on the SUSY mass scale. Hence the precision of 
this prediction is better than the current experimental precision and
matches the foreseen precision after LHC and a linear $e^+e^-$ collider.

\begin{figure}
\includegraphics[width=9.0cm]{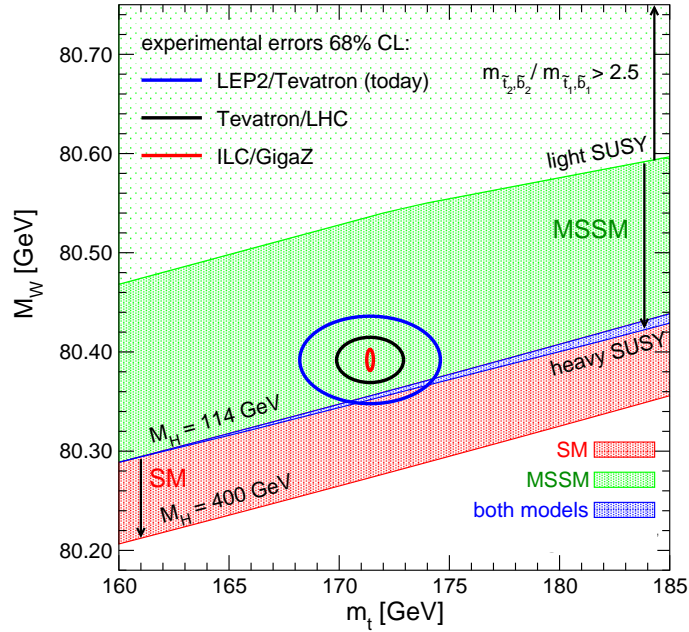}
\setlength{\unitlength}{1pt}
\begin{picture}(0,0)
\CBox(-110,30)(-18,40){White}{White}
\end{picture}
\caption{$M_W$ in the SM (red/medium shaded) and the MSSM (green/light shaded)
 as function  of the top mass
  $m_t$. The SUSY parameters are varied independently in a random
  parameter scan in the following ranges: the diagonal entries of the squark 
and slepton mass matrices are varied between $100\ldots2000$~GeV,  
$A_{t,b}$, $\mu=-2000\ldots2000$~GeV, $M_A=90\ldots1000$~GeV.
The SM prediction for
  $M_W$ is shown for $M_H=114\ldots400$~GeV.
\label{fig:MWMT}}
\end{figure}

The predictions for $M_W$ in the SM and the MSSM are compared in
Figure \ref{fig:MWMT}.
The possible predictions within the two models as a function of all model
parameters
give rise to two bands in the $m_t$--$M_W$ plane with only a relatively small
overlap region (blue area). For the employed parameter regions
see the caption of Figure \ref{fig:MWMT}.
The MSSM band is divided into two regions. 
In the very light-shaded green region
at least one of the ratios $m_{\tilde{t}_2}/m_{\tilde{t}_1}$ or 
$m_{\tilde{b}_2}/m_{\tilde{b}_1}$ exceeds~2.5, while in the green region
the mass ratios are unconstrained.

The current 68\%~C.L.\ experimental results
for $m_t$ and $M_W$ slightly favour the MSSM over the SM. 
More importantly, both within the MSSM and the SM, the precision
of the experimental measurements 
excludes large regions of parameter space.

The prospective accuracies for the LHC and the ILC with GigaZ option 
are also shown in the plot (using the current
central values), indicating the
potential for a significant improvement of the sensitivity of the
electroweak precision tests.

\section{$a_\mu$ in the MSSM}

The impressive measurement of the anomalous magnetic moment of the muon $a_\mu$
\cite{BNL56} has inspired a lot of progress on the theoretical side.
After many refinements (see \cite{Passera,DM} for recent reviews), the
SM prediction deviates by about two standard deviations from the final
experimental value, $\Delta a_\mu(\mbox{exp}-\mbox{SM}) = 23.9\,(9.9)
\times10^{-10}$ \cite{DM}. 

The MSSM prediction of $a_\mu$ has been reviewed recently in \cite{rev}.
The leading 
contributions from SUSY particles are approximately given by
$13\times10^{-10}\left({100\,\rm GeV}/{M_{\rm SUSY}}\right) ^2\times
\ \tan\beta\ \mbox{sign}(\mu)$ \cite{Lopez1L,Chatt1L,Moroi1L}. 
Due to the enhancement $\propto\tan\beta$,
the ratio
of the two Higgs vacuum expectation values, the SUSY contributions
could easily be the origin of the deviation 
$\Delta a_\mu(\mbox{exp}-\mbox{SM})$.
Furthermore, the positive value of this deviation implies a
preference for a positive $\mu$-parameter in the MSSM.

The status of the MSSM prediction is as follows. The one-loop contributions
have been known for a long time (e.g.\ \cite{Lopez1L,Chatt1L,Moroi1L};
for further references see \cite{rev}). 
The two-loop contributions can be devided
into two classes. The class with closed loops of SUSY particles, 
such as squark or chargino/neutralino loops, is completely known 
\cite{HSW03,HSW04}; the class without such loops is known in the leading-log 
approximation \cite{DG98new}. The one-loop and leading two-loop contributions
have a very compact analytical form, see \cite{rev} and references therein.
The remaining theoretical uncertainty due
to unknown higher-order contributions has been estimated to be smaller than
$3\times10^{-10}$ \cite{rev}. This is satisfactory at the moment, but it 
could be significantly improved by a computation of the remaining two-loop
contributions.

\begin{figure}[tb!]
\epsfbox{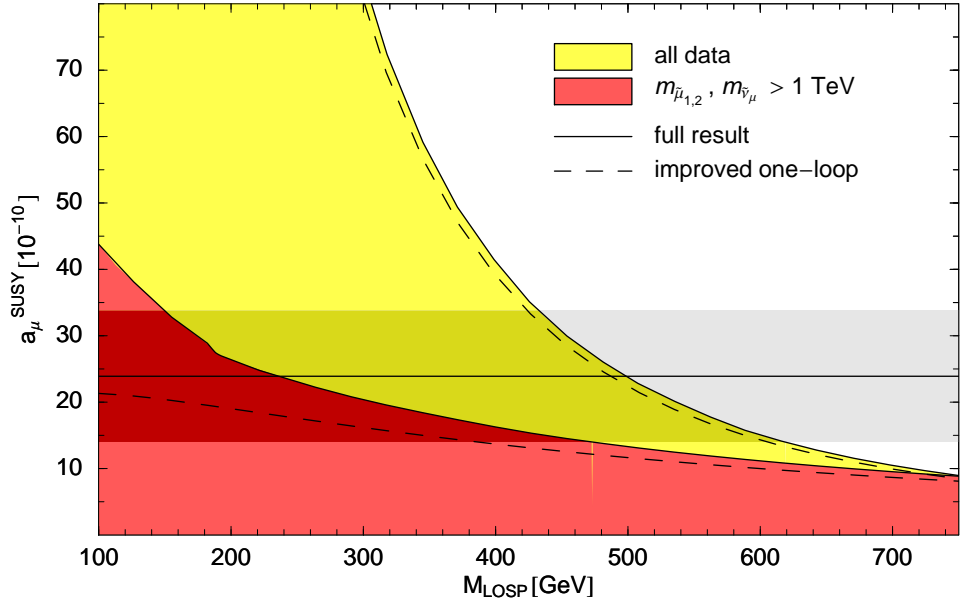}
\caption{\label{fig:amu}
Allowed values of MSSM contributions to $a_\mu$ 
as a function of the mass of the lightest observable SUSY 
particle $M_{\rm
  LOSP}=$min$(m_{\tilde{\chi}^\pm_1}, m_{\tilde{\chi}^0_2},
m_{\tilde{f}_i})$, from an MSSM parameter scan with $\tan\beta=50$ (see
\cite{rev} for the employed parameter ranges). The $1\sigma$ region
corresponding to the deviation between experimental and SM values
 is indicated. The light yellow region corresponds to all
input parameter
points that satisfy the experimental constraints from $b$-decays,
$M_h$ and $\Delta\rho$. In the red region, smuons and sneutrinos
are heavier 
than 1 TeV. The dashed lines correspond to the contours that arise
from ignoring the two-loop corrections from chargino/neutralino- and
sfermion-loop diagrams.
}
\end{figure}

The current status of $a_\mu$ in the MSSM is summarized in Figure
\ref{fig:amu}, which shows the possible MSSM contributions to $a_\mu$
compared with the observed deviation between experiment and the SM
prediction. Clearly, the MSSM can accomodate the experimental result for
$a_\mu$, and the preferred mass scale is rather low.

\end{document}